\documentclass[aps,prl,twocolumn,amssymb,superscriptaddress]{revtex4}
\usepackage{graphicx}
\usepackage{amsmath}
\usepackage{amssymb}
\usepackage{bm}

\begin{document}

 {\noindent \bf  Comment on \textquotedblleft Aharonov-Casher and Scalar Aharonov-Bohm Topological Effects \textquotedblright}

\vspace*{0.5\baselineskip}

 In a recent Letter \cite{Ma},
 Dulat and Ma drive
 (i) the relativistic Hamiltonian, Eq. (17), of a fixed spin
 from the spin-state projected Lagrangian $\mathcal{L}_{+}$, Eq. (6),
 with the $U(1)_{mm}$ gauge structure
 for a neutral spin half particle with an anomalous magnetic moment $\mu$,
 discuss
 (ii) the conditions, Eqs. (23)-(25), for a topological Aharonov-Casher (AC)
  and scalar Aharonov-Bohm (SAB) effects,
 and then
 make (iii) the conclusion that the arguments of Peshkin and Lipkin \cite{Peshkin},
 which state that the AC and SAB effects are not topological,
 are incorrect,
 by stating that ``they (Peshkin and Lipkin)
 are based on the wrong Hamiltonian which yields their incorrect conclusion."
 based on the observation
 that there is no $U(1)_{mm}$ gauge structure
 in the Hamiltonian of Peshkin and Lipkin.
 In this Comment we point out
 (i) that their Hamiltonian, Eq. (17), is not a relativistic Hamiltonian,
 (ii) then that the conditions are irrelevant for a topological AC and SAB effects,
 and (iii) conclusively that the non-relativistic Hamiltonian
 employed by Peshkin and Lipkin \cite{Peshkin}
 has the same $U(1)_{mm}$ gauge structure for a fixed spin
 and then is not wrong, but their incorrect interpretation of the spin autocorrelations
 led to the incorrect conclusion.

 In the Letter \cite{Ma},
 in order to obtain the underlying $U(1)_{mm}$ gauge symmetry
 in the Largrangian $\mathcal{L}$, Eq. (1),
 Dulat and Ma use the spin projection operator
 such that the polarization direction of the neutral
 particle does not vary.
 However, the spin-state projected Lagrangian $\mathcal{L}_+$, Eq. (6),
 cannot preserve the relativistic invariance
 because the spin should undergo a Wigner rotation under a general Lorentz transformation.
 This implies that a fixed spin is possible only in a certain reference frame.
 Consequently,
 the Hamiltonian, Eq. (17), driven from
 the Lagrangian ${\cal L}_+$, Eq. (6), is not a relativistic Hamiltonian.
 Subsequently, its non-relativistic approximated Hamiltonian,
 Eq. (18), is unphysical.
 For instance, for the AC interaction,
 the right non-relativistic Hamiltonian
 is in Eq. (21) driven by Aharonov and Casher in \cite{Casher}.
 Evidently, thus, the conditions, Eqs. (23)-(25),
 in order to recover the original AC
 or/and SAB Hamiltonians in their Hamiltonian, Eq. (18),
 are not physically meaningful for the necessary conditions of the AC and SAB setups.

 The total non-relativistic Hamiltonian of Peshkin and Lipkin, including both the AC and SAB effects,
 considered in \cite{Peshkin} can be actually written as
\begin{eqnarray}
  \mathcal{H} = \frac{1}{2m}
  \left( {\mathbf p} - \mu\, {\boldsymbol{\sigma}}\times {\mathbf E}\right)^2
  - \mu\, {\boldsymbol{\sigma}} \cdot{\bf B}
  \label{Heq}
\end{eqnarray}
 for a neutral particle with velocity ${\mathbf v}$
 in electric and magnetic fields ${\mathbf E}$ and ${\mathbf B}$.
 This Hamiltonian has the $SU(2)_{\mathrm spin}$ gauge symmetry.
 However, for fixed spins,
 Eq. (\ref{Heq}) has the same $U(1)_{mm}$ gauge structure with
 $\mathcal{A} = {\mathbf s} \times {\mathbf E} $
 and ${\mathcal A}_0 = {\mathbf s}\cdot {\mathbf B}$ as introduced in \cite{Ma}, where
 new electric and magnetic fields can be defined as
 ${\mathcal E}= -{\boldsymbol{\nabla}} {\mathcal A}_0$ and ${\mathcal B}={\boldsymbol{\nabla}} \times {\mathcal A}$, respectively.
 Then, $\mathcal{A}$ and $\mathcal{A}_0$ can induce the topological AC and SAB effects for fixed spins,
 respectively.
 To generate topological AC and SAB phases, as is known,
 the prerequisite conditions of force-free and torque-free are required for the AC and SAB setups.
 For a given ${\mathcal H}$, the force
 $\frac{d\mathbf p}{dt}=\left[ {\mathbf p}, {\mathcal H}\right]$
 and the spin torque $\frac{d\boldsymbol \sigma}{dt}=\left[ {\boldsymbol \sigma}, {\mathcal H}\right]$
 on the neutral particle should be zero for the topological AC and SAB setups.
 For the SAB setup, as an example,
 the force and the spin torque are given by
 $\frac{d\mathbf p}{dt}
 =i\hbar \mu {\boldsymbol \nabla} \left( {\boldsymbol \sigma} \cdot {\mathbf B}\right) $
 and $\frac{d\boldsymbol \sigma}{dt}
 =\mu\, \mbox{\boldmath$\sigma$} \times {\mathbf B}$
(introduced as Eq. (8) in \cite{Peshkin} and Eq. (27) in \cite{Ma}),
 respectively.
 Let us consider that the spin is fixed in the $\left|+z\right\rangle $ state
 and $+z$ is the direction of the magnetic field ${\mathbf B}=B_z \hat z$.
 Then, the force becomes zero,
 ${\boldsymbol \nabla} \left( {\mathbf s} \cdot {\mathbf B}\right) % ={\boldsymbol \nabla} {\mathcal A}_0
 =0 = {\mathcal E}$.
 The spin torque also becomes zero, as shown as follows:
 $\sigma_x$ and $\sigma_y$ are actually
 the spin flip operators so that the expectation value of the spin torque,
 $\left\langle +z(t)\right|\sigma_{x/y} B_z\left|+z(t) \right\rangle$,
 should be zero for fixed spins.
 Thus,
 no Larmor-type precession occurs, which
 cannot give the frequency $\omega = 2\mu B/\hbar$ in the spin correlations of Eq. (11) in
 \cite{Peshkin}.
% For fixed spins,
 Resultantly, Peshkin and Lipkin have missed
 the trivial solution of the spin correlation equations,
 $C(t)=0$ and $S(t)=0$ in Eq. (12) in \cite{Peshkin}.
 Indeed, Dulat and Ma have also noticed \cite{Ma} that
 the effects of the quantum fluctuations in Eq. (12) in \cite{Peshkin} can exist
 only at a very short time, $\Delta t \approx 0$.
 Consequently, the SAB and AC effects can be topological for fixed spins.
 The reason why Peshkin and Lipkin have made the incorrect conclusion in \cite{Peshkin}
 is not because of using a wrong Hamiltonian as Dulat and Ma have claimed in \cite{Ma},
 but because of their incorrect interpretation of the spin torque equation.

 We acknowledge support from the National Research
 Foundation of Korea Grant funded by the Korean
 Government(2012-0003786) (T.C.) and
 the National Natural Science Foundation of China under the Grant No. 11374379 (S.Y.C.).

\vspace*{0.5\baselineskip}

\noindent Taeseung Choi$^{1*}$ and Sam Young Cho$^{2\dagger}$

\noindent $^1${\small Division of General Education, Seoul Women's University,
    Seoul 139-774, Korea}

\noindent $^2${\small Center for Modern Physics and Department of Physics,
     Chongqing University, Chongqing 400030, China}

{\small
\indent $^*$Electronic address: tchoi@swu.ac.kr \\
\indent $^\dagger$Electronic address: sycho@cqu.edu.cn}

\end{document}